\documentclass[twocolumn,showpacs,showkeys,preprintnumbers,amsmath,amssymb,prd]{revtex4}
\usepackage{graphicx}
\usepackage{dcolumn}
\usepackage{bm}
\renewcommand{\vec}[1]{{\boldsymbol{#1}}}

\begin{document}

\title{On the kinetic equation approach to pair production\\ by time-dependent electric field}

\author{A.~M. Fedotov}
\email{fedotov@cea.ru}
\author{E.~G. Gelfer}
\affiliation{National Research Nuclear University MEPhI, Moscow 115409, Russia}
\author{K.~Yu. Korolev}
\affiliation{Russian Research Centre ``Kurchatov Institute'', Moscow 123182, Russia}
\author{S.~A. Smolyansky}
\affiliation{Saratov State University, Saratov 410026, Russia}

\begin{abstract}
We investigate the quantum kinetic approach to pair production from vacuum by time-dependent electric field. Equivalence between this approach and the more familiar S-matrix approach is explicitly established for both scalar and fermion cases. For the particular case of a constant electric field exact solution for kinetic equations is provided and the accuracy of low-density approximation is estimated.
\end{abstract}


\maketitle

\section{\label{sec:intro}Introduction}

It often occurs that a physical problem can be treated by means of rather different mathematical approaches. Problem of pair creation from vacuum by electric field was first considered on the basis of the first quantized relativistic theory in the context of the Klein paradox, i.e. scattering at electrical potential barrier \cite{Klein,Sauter,Somm}. Later on, this problem was studied in the framework of quantum field theory by the effective action technique \cite{HE,Weiss,Schw}, by explicit construction of $S$-matrix in the non-stationary gauge \cite{NN,Nik,Nik1,rev4} and by semiclassical methods \cite{BI,Popov}, see also the monographes \cite{rev1,rev2,rev3} for review of these and some other approaches.

More recently, there appeared yet another approach to pair creation effect in external field \cite{Rau,Smol,Kluger,Smol1,Smol2}, which is based on equations that bear strong resemblance to kinetic equations widely used in plasma physics and other non-equilibrium problems. Following the authors of these papers, let us call it the quantum kinetic equation (QKE) approach.
The obvious benefits of the QKEs are simple implementation of numeric calculations \cite{Smol,Smol1,Smol2} and ability of natural inclusion of account for back reaction to the external field \cite{Bloch}. On the other hand, there are also several disadvantages, including (at least, currently) restriction to homogeneous fields only and seemingly non-gauge invariant form, non-ability of finding exact analytical solutions even for the problems that knowingly admit such solutions, and, as we are demonstrating in this paper, delusive physical interpretation of the basic quantities involved in QKE. Obviously, synthesis of different approaches may be very helpful for further development of the theory.

The QKEs had been rigourously derived from the first principles of QED for a specific case of homogeneous time-variable electric field for creation of both scalar \cite{Kluger,Smol2} and fermion \cite{Smol,Smol1} pairs. Therefore, this approach must be manifestly equivalent to other exact methods mentioned above. Surprisingly, to the best of our knowledge, no attempts were made in the literature to establish this equivalence explicitly until the recent paper \cite{Dumlu}. Moreover, let us mention the attempts to contrast the results obtained by QKE to those obtained by more traditional approaches \cite{Gies} with finding difference among them, and to derive the different source term for QKE starting from the $S$-matrix approach \cite{Tanji}. As for the paper \cite{Dumlu}, though it have stated for the first time the aforementioned equivalence, explicit correspondence was provided only for scalar pair production and, in addition, in the form which is not best suited for practical purposes (e.g., for testing the numerical routines with the exact solutions).

In order to simplify the QKE, several ad hoc approximations had been suggested, such as, e.g., the Markovian and the low-density approximation
\cite{Schmidt,Gies}. However, expectations that these approximations can be justified under some conditions are based on the analogies with the properties of more usual kinetic equations. An important point here is that the function satisfying the QKE does not actually possess the meaning of  particle distribution as long as particle creation process goes on. Rather, this function accounts for a ``quantum soup'' of real particles and vacuum fluctuations. These ingredients can not be separated in principle owing to the uncertainty relations.

In the present paper, after a brief review of the QKE approach in the section \ref{sec:1}, we provide the explicit bilinear ansatz that converts QKE to the (dimensionally reduced) Klein-Gordon or Dirac (Sec.~\ref{sec:2}) equation. Our ansatz  allows to adopt the known exact analytical solutions of the pair creation problem in the QKE context. In particular, we illustrate our correspondence in more details for the paradigmatic case of constant electric field. In the Sec.~\ref{sec:3} we explicitly demonstrate that the low-density approximation is not asymptotically exact in the weak field limit and estimate its accuracy. The discussion and concluding remarks are collected in the Sec.~\ref{sec:4}.

\section{QKE approach: illustration with 1D oscillator model}
\label{sec:1}

Derivation of QKEs in the context of pair creation problem is thoroughly discussed in the literature \cite{Smol,Smol1,Smol2,Dumlu}. However, for convenience of the reader and in order to introduce the notation, let us provide here a sketch of derivation of QKE in a more simple toy model problem - for parametric excitation of a 1D quantum oscillator with time varying frequency $\omega(t)$. The familiar Hamiltonian of the oscillator reads
$$
H=\frac{P^2}{2}+\frac{\omega^2(t)Q^2}{2},
$$
where $Q$ and $P$ are the coordinate and the momentum operators, obeying the ordinary commutation relations.

Consider the \emph{approximate} WKB solution for the oscillator equation\footnote{We are using the Heisenberg picture and the units $\hbar=c=1$.} $\ddot Q+\omega^2(t)Q=0$,
\begin{equation}\label{WKB}
\Phi(t)=\frac{1}{\sqrt{2\omega(t)}}e^{-i\Theta(t)},\quad \Theta(t)=\int\limits_0^t
\omega(t)\,dt,
\end{equation}
and the time-dependent operators $a(t)$, $a^\dagger(t)$, defined in such a way that the relations
\begin{eqnarray}
Q(t)=\Phi(t)a(t)+\Phi^*(t)a^\dagger(t),\nonumber\\
\quad P(t)=-i\omega(t)[\Phi(t)a(t)- \Phi^*(t)a^\dagger(t)],\label{a1}
\end{eqnarray}
after all are carried out \emph{exactly}.
These operators are obviously expressed in terms of the original operators $Q$ and $P$ via
\begin{equation}\label{a2}
a(t)=(\omega(t)Q+iP)\Phi^*(t),\quad
a^\dagger(t)=(\omega(t)Q-iP)\Phi(t),
\end{equation}
are \emph{time-dependent} and obey the commutation relation $[a,a^\dagger]=1$.

Next, by taking into account the equations of motion for the oscillator, we obtain:
\begin{equation}\label{a_eqs}
\dot a(t)=\frac{\dot\omega(t)}{2\omega(t)}e^{2i\Theta(t)}a^\dagger(t),\quad \dot a^\dagger(t)=\frac{\dot\omega(t)}{2\omega(t)}e^{-2i\Theta(t)}a(t).
\end{equation}
Thus, time dependence of the introduced operators is caused solely by variation of the oscillator frequency.

Now, consider the ``instant excitation number'' $N(t)=a^\dagger(t)a(t)$. By taking into account the Eqs.~(\ref{a_eqs}), we can write
\begin{equation}\label{dN}
\dot N(t)=\frac{\dot\omega(t)}{2\omega(t)}(V+V^\dagger),
\end{equation}
where we have denoted $V(t)=e^{-2i\Theta(t)}a^2(t)$. On the same grounds, the time derivative of $V(t)$ can be expressed in the form
\begin{equation}\label{dA}
\dot V(t)=-2i\omega(t)A+\frac{\dot\omega(t)}{2\omega(t)}(1+2N(t)).
\end{equation}
The averages ${\cal N}(t)$, ${\cal V}(t)$ of the operators $N(t)$ and $V(t)$ over the actual quantum state of the oscillator obviously obey the same equations (\ref{dN}) and (\ref{dA}).

Suppose that initially (i.e., at $t\to-\infty$), the frequency was constant and the oscillator was hosted in some stationary state $|n\rangle$. This means, in particular, that the operators $a$ and $a^\dagger$ had represented just the usual lowering and raising operators acting on the ladder of the stationary states. In particular, the ``anomalous average'' ${\cal V}(-\infty)$ is zero initially under our assumption. With such initial condition, the equation (\ref{dA}) can be integrated out with the result
\begin{equation}\label{A_ex}
{\cal V}(t)=\int\limits_{-\infty}^t
dt'\,\frac{\dot\omega(t')}{2\omega(t')}[1+2{\cal N}(t')]e^{-2i[\Theta(t)-\Theta(t')]}.
\end{equation}
The analogue of the QKE for our toy model
\begin{equation}\label{QKE_osc}
\dot {\cal N}(t)=\frac{\dot\omega(t)}{2\omega(t)}\int\limits_{-\infty}^t
dt'\,\frac{\dot\omega(t')}{\omega(t')}[1+2{\cal N}(t')]\cos{2(\Theta(t)-\Theta(t'))},
\end{equation}
can be now obtained by substitution of the Eq.~(\ref{A_ex}) into the Eq.~(\ref{dN}).

Let us demonstrate how the integro-differential Eq.~(\ref{QKE_osc}), or, equivalently, the system of differential equations (\ref{dN}), (\ref{dA}) can be converted back into the original oscillator equation. Let $\Psi(t)$ be an exact, positive frequency solution of the equation
\begin{equation}\label{Psi_osc}
\ddot\Psi+\omega^2(t)\Psi=0.
\end{equation}
Under our assumptions, the latter means that (up to a constant phase factor) $\Psi(t)\to\Phi(t)$ as $t\to-\infty$.

The position and the momentum operators of the oscillator can be expressed in terms of the aforementioned solution by
\begin{equation}\label{qp_Psi}
Q(t)=\Psi(t)\alpha+\Psi^*(t)\alpha^{\dagger},\quad P(t)=\dot \Psi(t)\alpha+
\dot \Psi^*(t)\alpha^\dagger,\end{equation}
where the operators $\alpha$, $\alpha^\dagger$
are now time-independent. These operators obey the usual commutation relations $[\alpha,\alpha^\dagger]=1$, provided that the solution $\Psi(t)$ is normalized by
\begin{equation}\label{normaliz}
\dot\Psi\Psi^*-\Psi\dot\Psi^*=-i,
\end{equation}
and they acquire the meaning of in-operators (i.e., define the ladder of the initial states). By averaging the operator equalities
\begin{eqnarray}
N(t)=\frac{\omega^2(t)Q^2+P^2-\omega(t)}{2\omega(t)},\nonumber\\ V(t)=
\frac{\omega^2(t)Q^2-P^2-i\omega(t)(QP+PQ)}{2\omega(t)},\nonumber
\end{eqnarray}
which follow from the Eqs.~(\ref{a2}) over the initial state $|n\rangle=(\alpha^\dagger)^n|0\rangle/\sqrt{n!}$ (where $\alpha|0\rangle=0$), we find that
\begin{equation}\label{N_expl}
{\cal N}(t)=n+\frac{\omega^2(t)|\Psi(t)|^2
+|\dot \Psi(t)|^2-\omega(t)}{2\omega(t)}(1+2n),
\end{equation}
and
\begin{equation}\label{A_expl}
{\cal V}(t)=\left\{\frac{\omega^2(t)|\Psi(t)|^2 -|\dot
\Psi(t)|^2}{2\omega(t)}-\frac{i}2\frac{d}{dt}|\Psi(t)|^2\right\}(1+2n).
\end{equation}
It is simple to prove by direct substitution that the formulas (\ref{N_expl}), (\ref{A_expl}) indeed define the exact solution of equations (\ref{dN}), (\ref{dA}) with initial conditions ${\cal N}(-\infty)=n$, ${\cal V}(-\infty)=0$, and hence, of equation (\ref{QKE_osc}), see also the Appendix \ref{App_A}.

\section{Scalar and fermion pair production}
\label{sec:2}

Consider a charged scalar field $\varphi$ affected by a linearly polarized time-dependent homogeneous electric field, directed along the $z$ axis. In this section, we assume for simplicity that the quantum field is initially in its vacuum state. After separation of variables, the Klein-Gordon equation
\begin{equation}\label{KG_eq}
  \left[D_\mu D^\mu+m^2\right]\varphi(t,\mathbf{r})=0,\quad D_\mu=\partial_\mu+ieA_\mu
\end{equation}
for the function $\varphi(t,\mathbf{r})=(2\pi)^{-3/2}\Psi_\mathbf{p}(t)e^{i\mathbf{p}\mathbf{r}}$
is reduced to the form (\ref{Psi_osc}) with
\begin{equation}\label{omega}
\omega(t)=\sqrt{\pi^2+\epsilon^2},\quad \pi=p_z+eA(t),\quad \epsilon^2=p_\perp^2+m^2.
\end{equation}
Although this problem deals in fact with a complex, rather than the real valued oscillator, nevertheless the QKE acquires the same form (\ref{QKE_osc}) as above \cite{Kluger,Smol2}. Thus, its solution must be also of the same form (\ref{N_expl}), (\ref{A_expl}) as for the real valued oscillator. In accordance with our initial conditions, we only have $n_\mathbf{p}=0$.

The actual difference between the two cases is only in the meaning of the quantities ${\cal N}$ and ${\cal V}$. Namely, if $a_\mathbf{p}$ and $b_\mathbf{p}$ are the destruction operators for particles (with charge $-e$) and antiparticles (with charge $e$), respectively, then ${\cal N}_\mathbf{p}=\langle \mbox{vac}| a_\mathbf{p}^\dagger a_\mathbf{p}|\mbox{vac}\rangle$ is the ``instant average number of particles'' (or pairs), whereas the ``anomalous average'' is now defined as ${\cal V}_\mathbf{p}=e^{-2i\Theta_\mathbf{p}(t)}\langle \mbox{vac}| a_\mathbf{p} b_\mathbf{p}|\mbox{vac}\rangle $. Note that these ``anomalous averages'' are also important in condensed matter problems, e.g. they participate in formulation of semiconductor Bloch equations \cite{Semicond}.

Let's now turn to the case of the fermion (spinor) field. The Dirac equation
\begin{equation}\label{dir_eq}
[i\gamma^\mu D_\mu-m]\psi(t,\mathbf{r})=0,
\end{equation}
after transition to its squared version $[D^\mu D_\mu+m^2-i(e/2)F_{\mu\nu}\gamma^\mu\gamma^\nu]\psi=0$, for the function of the form $\psi(t,\mathbf{r})=(2\pi)^{-3/2}\Psi_\mathbf{p}^{(\pm)}(t)
e^{i\mathbf{p}\mathbf{r}}u_{\pm}$ ($\gamma^0\gamma^3u_{\pm}=\pm u_{\pm}$) reads
\begin{equation}\label{dir_eq1}
\ddot\Psi^{(\pm)}+[\omega^2(t)\pm ie\dot A(t)]\Psi^{(\pm)}=0,
\end{equation}
where $\omega(t)$ is defined by the same expression (\ref{omega}).

The QKE for the fermion case differs in its form from the Eq.~(\ref{QKE_osc}) \cite{Smol,Smol1},
\begin{eqnarray}
\dot {\cal N}(t)=\frac{\epsilon e\dot A(t)}{2\omega^2(t)}\int\limits_{-\infty}^t
dt'\,\frac{\epsilon e\dot A(t')}{\omega^2(t')}[1-2{\cal N}(t')]\nonumber\\
\times\cos\left[2(\Theta(t)-\Theta(t'))\right],\label{QKE_ferm}
\end{eqnarray}
at the same time the analogues for (\ref{dN}), (\ref{dA}) read
\begin{equation}\label{dN_ferm}
\dot{\cal N}(t)=\frac{\epsilon e\dot A(t)}{2\omega^2(t)}({\cal V}+{\cal V}^*),
\end{equation}
\begin{equation}\label{dA_ferm}
\dot {\cal V}(t)=-2i\omega(t){\cal V}+\frac{\epsilon e\dot A(t)}{2\omega^2(t)}(1-2{\cal N}).
\end{equation}
The functions ${\cal N}$ and ${\cal V}$ are defined similarly as for the scalar field, but are carrying an additional two-valued spin index. We omit this index as well as the momentum index for simplicity, note that the coefficients of the QKE are independent on the spin index.

\begin{figure*}
\includegraphics[width=8.5cm]{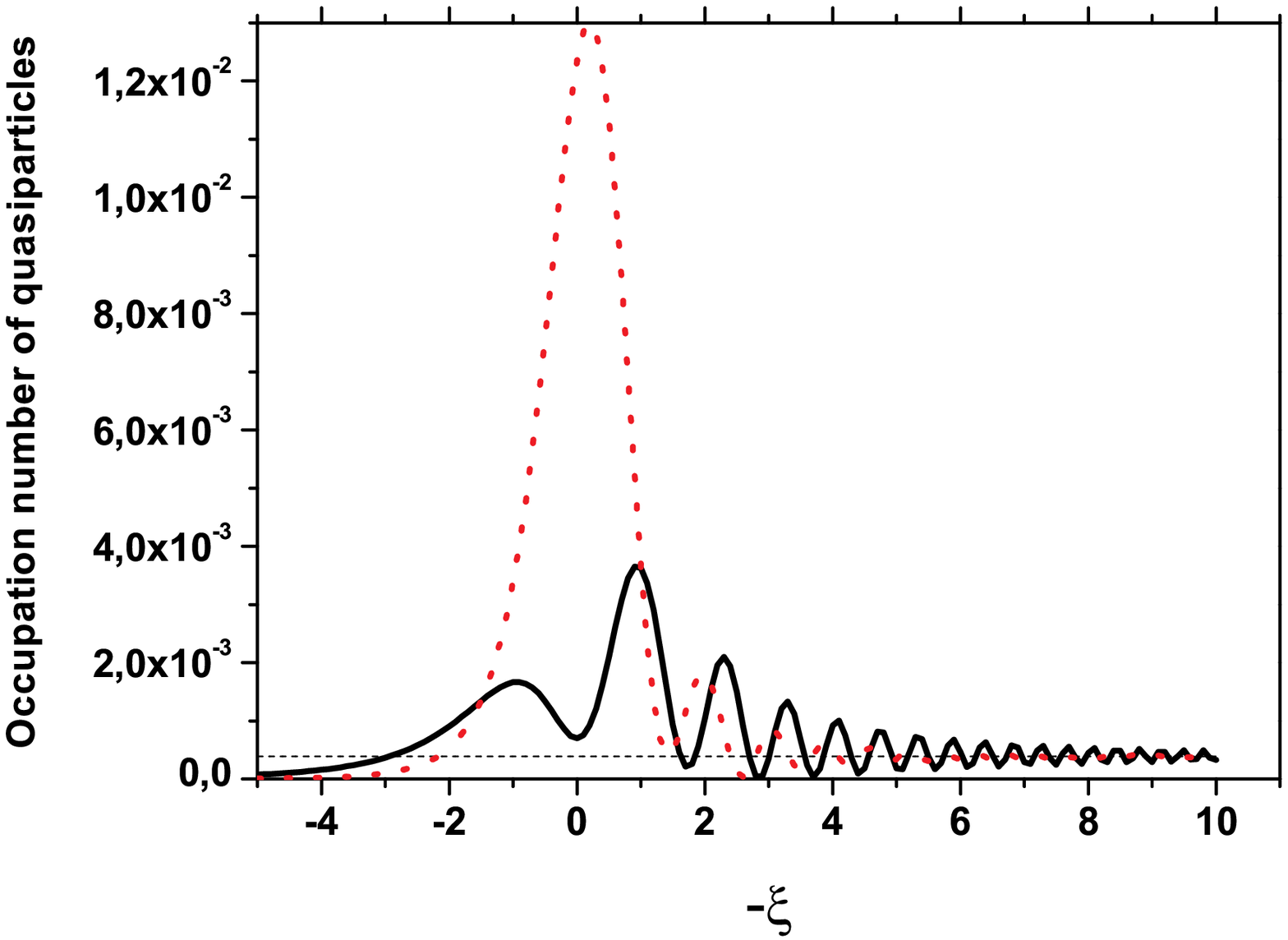}
\includegraphics[width=8.1cm]{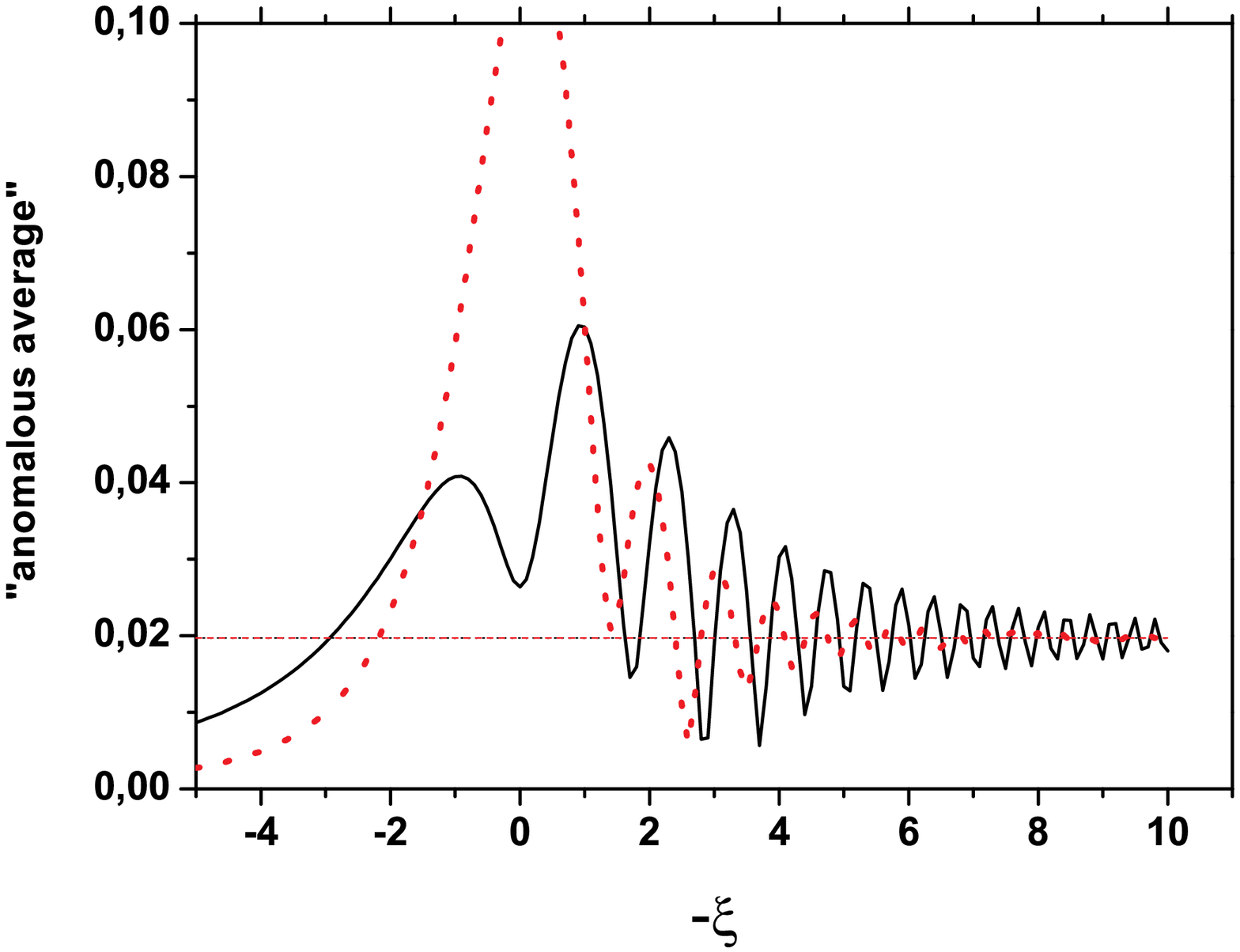}
\caption{\label{fig:lambda25} Occupation number ${\cal N}$ (left panel) and the magnitude of ``anomalous average'' ${\cal V}$ (right panel) for scalar (dotted line) and fermion (solid line) fields versus the dimensionless time $-\xi=\sqrt{eE}(t-p_z/eE)$ for $\lambda=2.5$. The dash lines show the asymptotic values at $t\to+\infty$.}
\end{figure*}

In the fermion case, reduction of the equations (\ref{dN_ferm}), (\ref{dA_ferm}) to the Eq.~(\ref{dir_eq1}) is less trivial than in scalar case, mostly due to a different ``statistical weight'' factor $1-2{\cal N}$ on the RHS. However, this procedure can still be done, e.g. following the derivation of the QKE. Let us present here only the result, which can be easily proved by direct substitution. Namely, we have
\begin{eqnarray}
{\cal N}(t)=\frac1{2\omega(\omega-\pi)}\big\{\omega^2|\Psi^{(\pm)}|^2
+|\dot \Psi^{(\pm)}|^2\nonumber\\\mp i\omega(\dot\Psi^{(\pm)}\Psi^{(\pm)*}-\Psi^{(\pm)}\dot\Psi^{(\pm)*})\big\},
\label{N_expl1}
\end{eqnarray}
\begin{eqnarray}
{\cal V}(t)=-\frac1{2\omega\epsilon}\big\{\omega^2|\Psi^{(\pm)}|^2
-|\dot \Psi^{(\pm)}|^2\nonumber\\\mp i\omega(\dot\Psi^{(\pm)}\Psi^{(\pm)*}+\Psi^{(\pm)}\dot\Psi^{(\pm)*})\big\},
\label{A_expl1}
\end{eqnarray}
where the functions $\Psi^{(\pm)}$ obey the Eq.~(\ref{dir_eq1}) with upper and lower sign, respectively, and are subject to the normalization condition
\begin{equation}\label{normaliz1}
\omega^2|\Psi^{(\pm)}|^2
+|\dot \Psi^{(\pm)}|^2\mp i\pi(\dot\Psi^{(\pm)}\Psi^{(\pm)*}-\Psi^{(\pm)}\dot\Psi^{(\pm)*})=\epsilon^2,
\end{equation}
see Appendix \ref{App_B} for more details. In practice, either of the functions $\Psi^{(\pm)}$ can be used to compute the quantities ${\cal N}$ and ${\cal V}$.

In the rest of the section, let us apply our results to the particular case of a constant electric field, $A(t)=-Et$. As it was stressed in the introductory section, though this problem is known to be solved exactly in terms of the parabolic cylinder functions $D_\nu$, no exact solution had been ever presented previously for QKEs. The positive frequency solutions for this case are explicitly given by (compare to \cite{NN,Nik,Nik1,rev4})
\begin{eqnarray}\label{scal_const_E}
\Psi_{scal}(t)=\frac{e^{-\pi\lambda/8}}{(2eE)^{1/4}}
D_{\frac{i\lambda-1}2}[(1-i)\xi],\\
\Psi_{ferm}^{(+)}(t)=e^{-\pi\lambda/8}D_{\frac{i\lambda}{2}}[(1-i)\xi],\nonumber\\
\Psi_{ferm}^{(-)}(t)=e^{-\pi\lambda/8}\sqrt{\frac{\lambda}{2}}
D_{\frac{i\lambda}{2}-1}[(1-i)\xi],\label{ferm_const_E}
\end{eqnarray}
where $\lambda=\epsilon^2/eE$ and $\xi=-\sqrt{eE}(t-p_z/eE)$. These formulas, together with the general expressions (\ref{N_expl}), (\ref{A_expl}) and (\ref{N_expl1}), (\ref{A_expl1}), define the desired solution for QKE in constant electric field.

\begin{figure*}[thhh]
\includegraphics[width=8.5cm]{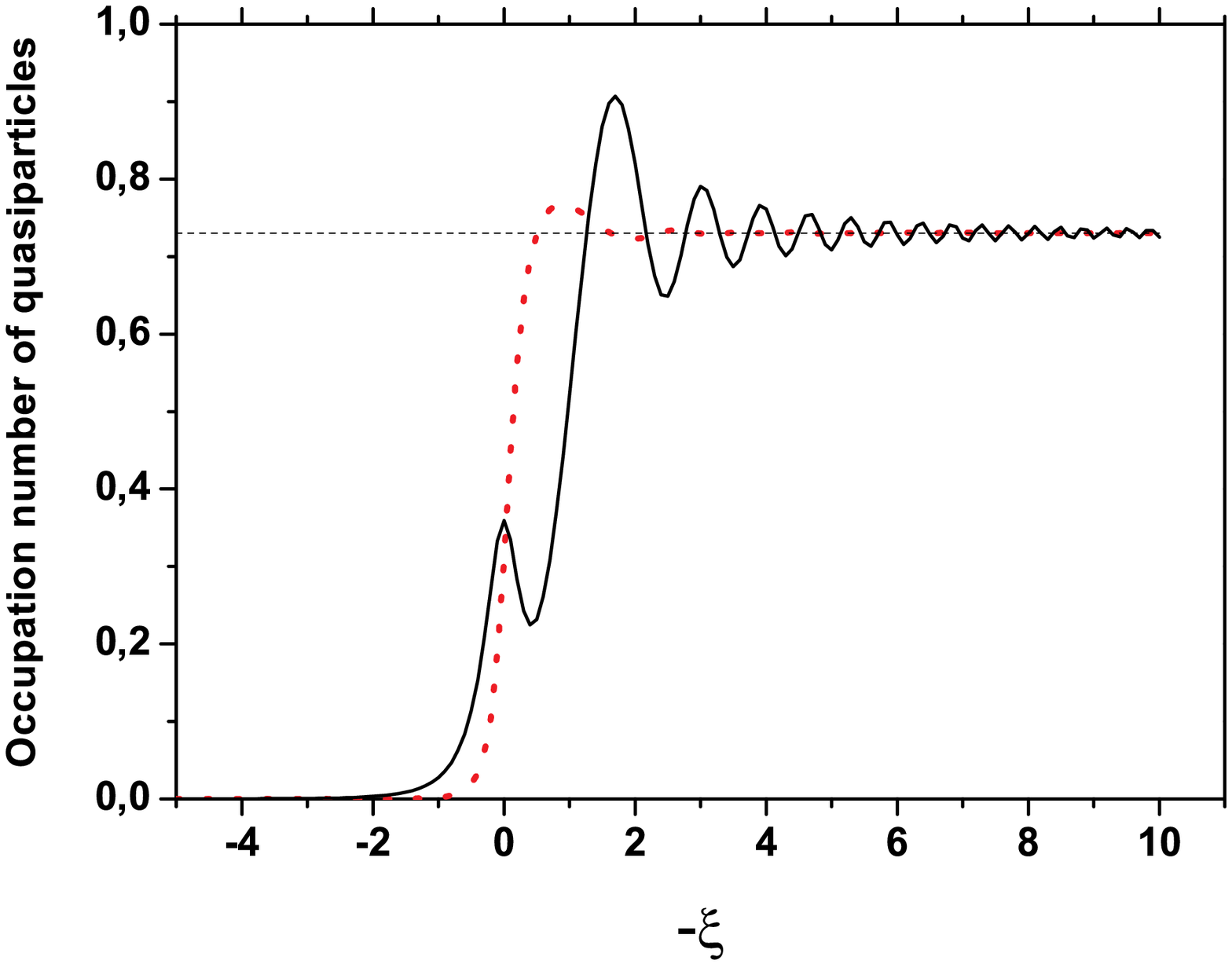}
\includegraphics[width=8.5cm]{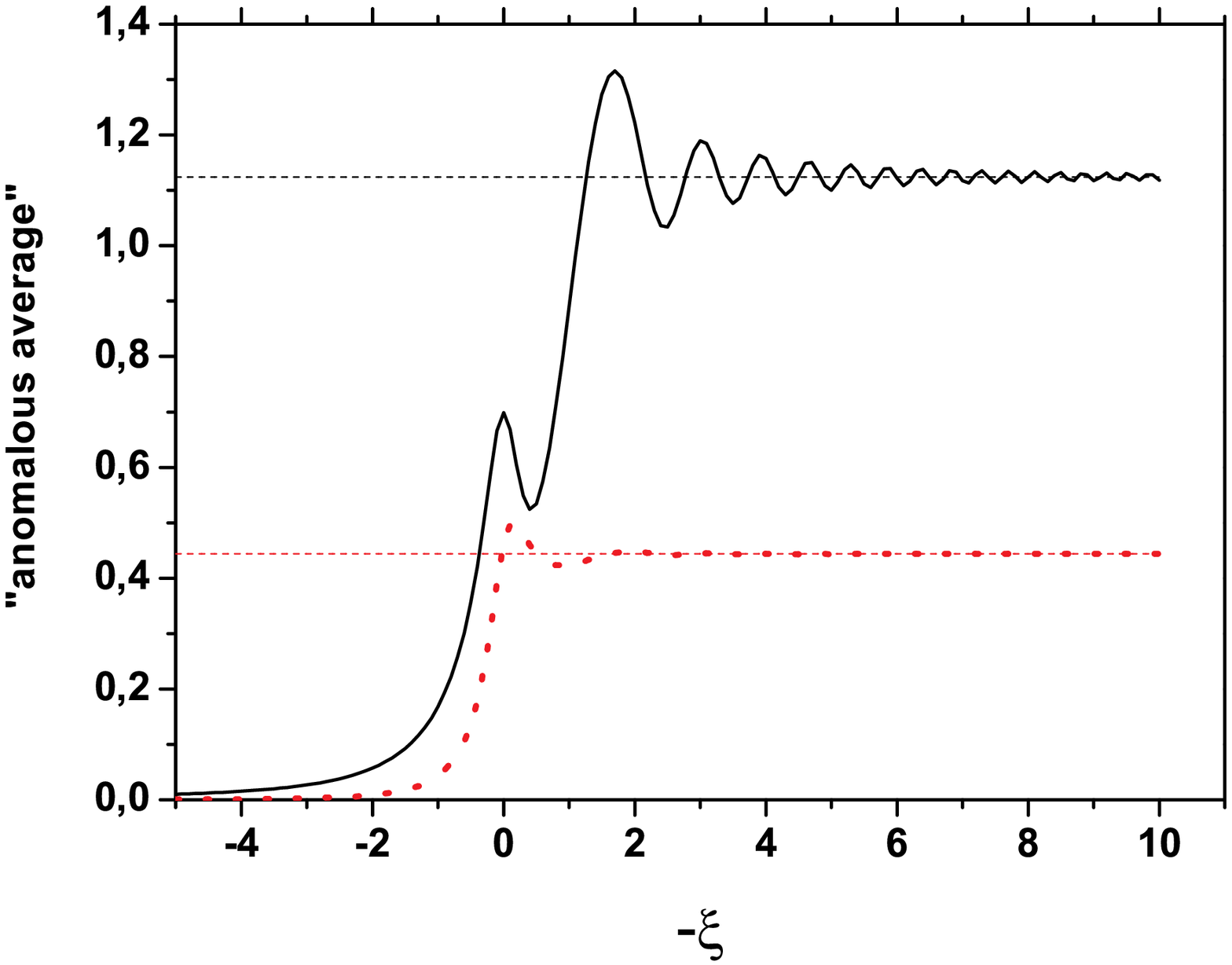}
\caption{\label{fig:lambda01} Occupation number ${\cal N}$ (left panel) and the magnitude of ``anomalous average'' ${\cal V}$ (right panel) for scalar (dotted line) and fermion (solid line) fields versus the dimensionless time $-\xi=\sqrt{eE}(t-p_z/eE)$ for $\lambda=0.1$. The dash lines show the asymptotic values at $t\to+\infty$.}
\end{figure*}

The dependence of the quantities ${\cal N}$ and $|{\cal V}|$ on time for both cases of scalar and fermion field is shown at Figs.~\ref{fig:lambda25} and \ref{fig:lambda01} for $\lambda=2.5$ (subcritical pair production) and $\lambda=0.1$ (supercritical pair production), respectively. The initial state is always assumed to be vacuum. The plots ${\cal N}(t)$ obtained by direct numerical integration of QKE can be found on Figs.~2 and 3 of the Ref.~\cite{Schmidt}. It is clear from the figures, that the curves are starting from zero at $t\to-\infty$ in accordance with our initial conditions and are tending to some asymptotic values at $t\to+\infty$. These asymptotic values are ${\cal N}(+\infty)=e^{-\pi\lambda}$ (for both scalars and fermions) and $|{\cal V}(+\infty)|=e^{-\pi\lambda/2}(1\pm e^{-\pi\lambda})^{1/2}$ (for bosons and fermions, respectively) and agree with the results obtained by other, e.g., $S$-matrix, methods.

However, this transition is not monotonous but is accompanied with oscillations in some transient region of width $\Delta t\sim \sqrt{\lambda/eE}$ around $t=p_z/eE$, which in fact essentially coincides to the commonly accepted ``coherence length'' (or formation time) \cite{NN,Nik,Nik1,rev4} for a pair production process\footnote{There is some discrepancy in the literature in definition of the coherence length. For example, in the Ref.~\cite{rev4} it is argued on different grounds that $\Delta t\sim {\rm max}\{1,\lambda\}/\sqrt{eE}$. Both estimations are compatible with our calculations. For the purposes of the present paper, this difference is not important.}. It can be also observed from the figures that the magnitude of oscillations is relatively much larger (compared to the corresponding asymptotic values) in the case of large values of $\lambda$, i.e. for subcritical pair production.
These oscillations can not be attributed to instant increase in particle production itself, because the notion of particles and antiparticles can not be defined rigorously in the transient region. This must be clear already from the $\dot N-N$ commutation relation
\begin{equation}\label{cr}
[\dot N,N]=\frac{\dot\omega}{\omega}\left(V-V^\dagger\right)
\end{equation}
which follows from the Eq.~(\ref{dN}). As a consequence, we have the uncertainty relation
\begin{equation}\label{ur}
\Delta\dot{\cal N}\cdot\Delta {\cal N}\ge\frac{|\dot\omega {\rm Im}{\cal V}|}{\omega}.
\end{equation}
Similar uncertainty relation can be derived for fermions. In virtue of (\ref{ur}), oscillations in the transient region are accompanied with quantum fluctuations. Similarly to ${\cal N}$, the quantity ${\rm Im}{\cal V}$ on the RHS also oscillates with frequency $2\omega$. The quantity ${\cal N}$ characterizes simultaneously the real particles and the vacuum perturbations and there is no way to extract information relevant for the particles alone. Hence, it is rather more accurate to use the term ``quasiparticles'' instead of ``particles'' in this context. Only outside the transient region, where $\omega$ varies slowly, quasiparticles turn into the real particles. This means that, in contradiction with the claims of some other authors, particle creation process can not be resolved on the temporal and spatial scales of the coherence length. We'll come back and give more comments on this peculiarity in the sequel.

The total number of pairs created in a unit volume during the time period $t_1<t<t_2$ can be obtained as usually by summing over all the excited modes of the field. For scalar field, this quantity can be computed as
\begin{eqnarray}
\int\limits\frac{d^3p}{(2\pi)^3}{\cal N}_{\bf p}(+\infty)=
\int\limits_{eEt_1}^{eEt_2}\frac{dp_z}{(2\pi)}\int\limits\frac{d^2p_\perp}{(2\pi)^2}
e^{-\pi\frac{m^2+p_\perp^2}{eE}}\nonumber\\=\frac{e^2E^2}{8\pi^3}e^{-\frac{\pi m^2}{eE}}(t_2-t_1),\label{N_tot}
\end{eqnarray}
where we have taken into account that only particles with $t_1<p_z/eE<t_2$ are being created in the time interval under consideration. For fermion case, the pair production yield is twice larger due to additional summation over the spin index of the modes. These results are well known, of course, and had been obtained previously by other methods, though in some modern papers the pair production yield is sometimes incorrectly identified with the doubled imaginary part of the Heisenberg-Euler-Schwinger effective action. The latter is related in fact to the vacuum-vacuum transition probability and has no exact relation to the pair production yield.

\section{Low-density approximation}
\label{sec:3}

The QKEs (\ref{QKE_osc}), (\ref{QKE_ferm}) look very complicated, in particular they possess non-Marcovian character. The latter means that the RHS of the QKE depends on the history. Of course, this dependence arises just for elimination of the equations (\ref{dA}), (\ref{dA_ferm}), so that the problem admits an equivalent formulation in local terms. Nevertheless, it may be useful in some problems to deal with QKE on its own footing, without reference to differential equations that lie in its origin. Several attempts were made in order to simplify the QKE, including the Markovian (taking the factor $1\pm 2{\cal N}$ outside the sign of the integral over $t'$) and the low-density (neglecting ${\cal N}$ on the RHS) approximations \cite{Schmidt,Gies}. Intuitively, they would be reasonable if ${\cal N}$ was large and monotonously increasing, or essentially small, respectively.

However, the occupation numbers are actually not large and, besides, the oscillatory behavior discussed in the preceding section generally contradicts both assumptions. In practice, it was shown by direct numerical integration in the papers cited above, that the Markovian approximation does not provide good accuracy (though, of course, gives the correct order) in the whole range of the field strength, whereas it is believed that the low-density approximation works good for weak fields. In this section, we evaluate the total pair production rate within the low-density approximation analytically for weak constant field and explicitly estimate its accuracy.

For definiteness, let us illustrate the line of computations using the Eq.~(\ref{QKE_osc}) and later come back to the fermion case. Within the low-density approximation, the total pair production rate per unit volume is defined by
\begin{equation}\label{low-dens}
R_{\mbox{ld}}=\int\limits\frac{d^3p}{(2\pi)^3}
\frac{\dot\omega_\vec{p}(t)}{2\omega_\vec{p}(t)}\int\limits_{-\infty}^t
dt'\,\frac{\dot\omega_\vec{p}(t')}{\omega_\vec{p}(t')}
\cos\left[2\int\limits_{t'}^t\omega_\vec{p}(t'')\,dt''\right]
\end{equation}
where we have recovered the momentum index of the modes. For the case of constant field, we have $\pi(t)=p_z-eEt$, so that we can change the variable of integration $p_z$ by $w=\pi(t)/\epsilon$. The time derivative $\dot\omega_\vec{p}(t)=eE\pi(t)/\omega_\vec{p}(t)$. Also, let us come to the new variables $u=\pi(t')/\epsilon=[\pi(t)+eE(t-t')]/\epsilon$ and $v=\pi(t'')/\epsilon=[\pi(t)+eE(t-t'')]/\epsilon$ instead of $t'$ and $t''$, respectively. After these transformations, we are coming to
\begin{eqnarray}
R_{\mbox{ld}}=\frac{eE}{2}\int\limits\frac{d^2p_\perp}{(2\pi)^3}
\int\limits_{-\infty}^{\infty} \frac{dw\,w}{1+w^2}
\int\limits_{w}^{\infty} \frac{du\,u}{1+u^2}\nonumber\\\times
\cos\left\{2\lambda[g(u)-g(w)]\right\},\label{low-dens1}
\end{eqnarray}
where $g(x)=\int_0^x\sqrt{1+v^2}\,dv$.
The integrand is invariant under exchange $u\leftrightarrow w$, therefore it is possible to represent the expression on the RHS in the form
\begin{eqnarray}
R_{\mbox{ld}}=\frac{eE}{4}\int\limits\frac{d^2p_\perp}{(2\pi)^3}
|f_{scal}(2\lambda)|^2,\nonumber\\ f_{scal}(x)=\int\limits_{-\infty}^{\infty} \frac{du\,u}{1+u^2}
e^{ixg(u)}.\label{low-dens2}
\end{eqnarray}
The same representation is valid in the fermion case as well, but with an additional factor two due to summation over the spin variables and with $f_{ferm}(x)=\int_{-\infty}^{\infty} du\,
e^{ixg(u)}/(1+u^2)$.

In the weak field limit $\lambda$ is large, so that the functions $f_{scal}$, $f_{ferm}$ can be evaluated by the approximate methods (see the Appendix \ref{App_C}). Thus we obtain
\begin{eqnarray}
f_{scal}(2\lambda)=\frac{2\pi i}{3} e^{-\pi\lambda/2}[1+O(\lambda^{-2/3})],\nonumber\\ f_{ferm}(2\lambda)=\frac{2\pi }{3} e^{-\pi\lambda/2}[1+O(\lambda^{-2/3})].
\label{f}
\end{eqnarray}
The integral in (\ref{low-dens2}) is evaluated easily and we discover that even for $\lambda\gg 1$ the total pair production rate in the low-density approximation exceeds the exact value given by (\ref{N_tot}) by the factor $(\pi/3)^2$, i.e. by $9.6\%$. We attribute this discrepancy to contribution of oscillations in the course of pair creation that have been discussed previously.

\section{Discussion}
\label{sec:4}

We have demonstrated that the quantum kinetic equation (QKE) approach to pair production problem is equivalent to the known for a long time $S$-matrix approach by establishing explicit correspondence between them for both scalar and fermion fields. This equivalence has been already declared recently \cite{Dumlu}, though not proved for the fermion case. As an illustration, we have applied our correspondence to obtain the exact solution for both scalar and fermion QKEs in the constant field. Of course, exact analytical solutions for QKEs can be now constructed for other situations that are known to be exactly solvable by conventional methods, e.g. for $E(t)=E_0/\cosh^2(\kappa t)$.

At the same time, any attempt to perturb the KQEs by applying some kind of approximation procedures (e.g., by imposing the Markovian, or the low-density approximations) inevitably destruct the correspondence. We have demonstrated it on the example of the low-density approximation which was believed to be well justified in the weak field limit. In contradiction with these expectations, we have proved that for the case of constant field its accuracy all the same tends to $10\%$ even under the optimal conditions.

It could seem that, being equivalent to the conventional approaches and looking more complicated, the QKEs are useless for pair creation problems with the possible exception of arranging the algorithms for numerical computations. This is not the case, however, since the QKE approach admits simple incorporation of backreaction on the external field \cite{Bloch}. In addition, the form of the kinetic equation is very pleasant in principle for incorporation of the secondary processes in the resulting electron-positron plasma \cite{Smol3,Smol4}, such as annihilation, hard photon emission, Compton scattering, etc. by simple inclusion of the appropriate collision terms on the RHS.

At the same time, in our opinion, at the present stage of development of QKE approach the latter attempts are too hasty, at least in the high-intensity laser problems. The first reason is that the spatial and temporal scales of variation of the laser field are of the same order, so that spatial variation of the field should be taken into account as well. Up to now, this has never been done starting from the first principles of QED, though some phenomenological conjectures had been given. The second reason is delusive physical interpretation of the occupation number participating QKEs at the time scales of the order of the formation time of pair creation process. As we have shown in the paper, the situation here is quite similar to the attempts to resolve the trajectory of a particle with the accuracy exceeding the limits imposed by the uncertainty principle. We believe that the true kinetic equations for pair creation, which can be treated by the commonly accepted methods, must contain some sort of averaging over the fine scale of formation for the pair creation process.

\begin{acknowledgments}
We are grateful to A.~A. Volodin from CMC MSU for assistance with evaluation of the integrals in the Appendix \ref{App_C}. We acknowledge partial support of this work from the grants RFBR~09-02-01201-a, RFBR~09-02-12201-ofi$\_$m and RNP~2.1.1/1871.
\end{acknowledgments}

\appendix

\section{Direct proof of the ansatz}
\label{App_A}
Let us show the way how our solution (\ref{N_expl}), (\ref{A_expl}) can be  deduced directly from the Eqs.~(\ref{dN}), (\ref{dA}), without any reference to derivation of the QKE. For this purpose, let us introduce the notation ${\cal V}={\cal R}+i{\cal J}$ for the real and imaginary parts of the ``anomalous average'' ${\cal V}$. In this notation, the equations (\ref{dN}) and (\ref{dA}) take the form
\begin{eqnarray}
\dot {\cal N}=\frac{\dot\omega}{\omega}{\cal R},\label{eq_n}\\ \dot {\cal R}=-2\omega {\cal J}+\frac{\dot\omega}{\omega}\left({\cal N}+\frac12\right),\label{eq_r}\\\dot {\cal J}=2\omega {\cal R}.\label{eq_j}
\end{eqnarray}
Let us consider a sum of Eqs.~(\ref{eq_n}), (\ref{eq_r}) and introduce the new function $\Upsilon$ defined by ${\cal N}+{\cal R}+1/2=\Upsilon\omega$. Then we immediately get
\begin{equation}\label{J_Ups}
{\cal J}=-\dot \Upsilon/2
 \end{equation}
and after that in virtue of Eq.~(\ref{eq_j})
\begin{equation}\label{R_Ups}
{\cal R}=-\ddot \Upsilon/4\omega,
\end{equation}
so that
\begin{equation}\label{N_Ups}
{\cal N}=\omega \Upsilon+\ddot \Upsilon/4\omega-1/2.
\end{equation}
Thus, all the functions participating  the system of equations (\ref{eq_n}), (\ref{eq_r}), (\ref{eq_j}) are now expressed in terms of a single unknown function $\Upsilon$.

Combining all the results (\ref{J_Ups}), (\ref{R_Ups}) and (\ref{N_Ups}) in either Eq.~(\ref{eq_n}) or Eq.~(\ref{eq_r}), we are coming to the third order equation for $\Upsilon$: $\Upsilon'''=-4\omega (\omega \Upsilon)'$. Multiplying it by $\Upsilon$ and integrating by taking account of the initial conditions $\Upsilon(-\infty)=(n+1/2)/\omega(-\infty)$, $\dot \Upsilon(-\infty)=\ddot \Upsilon(-\infty)=0$ and the identity $\Upsilon\Upsilon'''=(\Upsilon\Upsilon'')'-\Upsilon'\Upsilon''$,
we are arriving to
$$
\Upsilon\ddot \Upsilon-\frac{\dot \Upsilon^2}{2}+2\omega^2\Upsilon^2=\frac{(1+2n)^2}2.
$$
Now, substituting
\begin{equation}\label{Ups_Psi}
\Upsilon=|\Psi|^2(1+2n),
\end{equation}
after simple transformations we are finally coming to
\begin{eqnarray}
|\Psi|^2\left\{(\ddot \Psi+\omega^2 \Psi)\Psi^*+(\ddot \Psi^*+\omega^2
\Psi^*)\Psi\right\}\nonumber\\ = \frac{1+(\dot \Psi \Psi^*-\dot\Psi^*\Psi)^2}{2}.\label{Psi_sq}
\end{eqnarray}
Since the derivative of the RHS of Eq.~(\ref{Psi_sq}) can be represented in the form
$$
(\dot \Psi \Psi^*-\dot\Psi^*\Psi)\times[(\ddot \Psi+\omega^2 \Psi)\Psi^*-(\ddot \Psi^*+\omega^2\Psi^*)\Psi],
$$
it is clear that the sought-for function $\Psi$ satisfies the oscillator equation (\ref{Psi_osc}) together with the normalization condition\footnote{The as well admissible opposite sign on the RHS of this normalization condition is equivalent to exchange in notation $\Psi\leftrightarrow\Psi^*$.} (\ref{normaliz}), i.e. is a positive frequency solution at $t\to-\infty$. Finally, substituting (\ref{Ups_Psi}) into the Eqs.~(\ref{J_Ups}), (\ref{R_Ups}) and (\ref{N_Ups}) and using (\ref{Psi_osc}) to eliminate the second-order derivatives, we obtain the desired expressions (\ref{N_expl}) and (\ref{A_expl}).

\section{Normalization condition for fermion problem}
\label{App_B}

In order to derive the condition (\ref{normaliz1}), let us first multiply the Eq.~(\ref{dir_eq1}) and its conjugate by $\dot\Psi^{(\pm)*}$ and $\dot\Psi^{(\pm)}$, respectively, and take a sum of both of them. The resulting equation can be written in the form
\begin{eqnarray}
\left(\omega^2|\Psi^{(\pm)}|^2
+|\dot \Psi^{(\pm)}|^2\right)'=2\omega\dot\omega |\Psi^{(\pm)}|^2\nonumber\\\pm
i\dot\pi(\dot\Psi^{(\pm)}\Psi^{(\pm)*}-\Psi^{(\pm)}\dot\Psi^{(\pm)*}).
\label{norm_der1}
\end{eqnarray}

Next, multiplying the Eq.~(\ref{dir_eq1}) and its conjugate by $\Psi^{(\pm)*}$ and $\Psi^{(\pm)}$, respectively, and taking their difference, we obtain
\begin{equation}\label{norm_der2}
(\dot\Psi^{(\pm)}\Psi^{(\pm)*}-\Psi^{(\pm)}\dot\Psi^{(\pm)*})'\pm2i\dot\pi
|\Psi^{(\pm)}|^2=0.
\end{equation}
Taking into account that $\dot\omega=\pi\dot\pi/\omega$ and expressing $2\dot\pi |\Psi^{(\pm)}|^2$ from (\ref{norm_der2}), we can integrate the Eq.~(\ref{norm_der1}) with the result
\begin{equation}\label{norm_der3}
\omega^2|\Psi^{(\pm)}|^2
+|\dot \Psi^{(\pm)}|^2\mp i\pi(\dot\Psi^{(\pm)}\Psi^{(\pm)*}-\Psi^{(\pm)}\dot\Psi^{(\pm)*})={\rm const}.
\end{equation}
The constant on the RHS can be evaluated in the region where $\omega(t)$ varies slowly so that WKB approximation can be applied to the Eq.~(\ref{dir_eq1}). For this purpose, we restore the Planck constant in this equation as follows,
\begin{equation}\label{dir_eq1_h}
\hbar^2\ddot\Psi^{(\pm)}+[\omega^2(t)\pm i\hbar\dot \pi(t)]\Psi^{(\pm)}=0,
\end{equation}
and seek for the solution in the form $\Psi^{(\pm)}=e^{iS/\hbar}$, where $S=S_0+\hbar S_1+O(\hbar^2)$. Substituting this ansatz into Eq.~(\ref{dir_eq1_h}) and equating the coefficients at the same powers of $\hbar$, after all we find the WKB positive frequency solution in the form
\begin{equation}\label{WKB_sol}
\Psi^{(\pm)}\backsimeq C_\pm \left(\frac{\omega(t)\pm\pi(t)}{\omega(t)}\right)^{1/2}e^{-\frac{i}{\hbar}\int\omega(t) dt}.
\end{equation}
Being the components of a unique spinor, $\Psi^{(\pm)}$ are normalized by
$|\Psi^{(+)}|^2+|\Psi^{(-)}|^2=1$, so that $C_\pm=1/\sqrt{2}$. Thus, assuming $\hbar=1$, substituting (\ref{WKB_sol}) into the LHS of the condition (\ref{norm_der3}) and taking into account that in the framework of WKB approximation $\dot\Psi^{(\pm)}\backsimeq -i\omega\Psi^{(\pm)}$, we are coming to the Eq.~(\ref{normaliz1}).

\section{Calculation of the functions $f_{scal}$ and $f_{ferm}$}
\label{App_C}

Both functions are calculated in a similar manner. Let us start by evaluating the function $f_{ferm}(x)=\int_{-\infty}^{\infty} du\,F(u,x)$, $F(u,x)=e^{ixg(u)}/(1+u^2)$, assuming that $x$ is large. The stationary phase method can not be applied in our case, because the integral has poles at the stationary points $u=\pm i$ of the exponent and, in addition, these points are the branch points for $g'(u)$ rather than simple zeros.

Since the function $g(u)$ is odd, we can write
$$
f_{ferm}(x)=2\,{\rm Re}\,\int\limits_0^{\infty} du\,\frac{e^{ixg(u)}}{1+u^2}.
$$
Next, we have ${\rm Im}\,g(u)\to+\infty$ as $|u|\to\infty$ in the sector $0<\arg{u}<\pi/2$. Thus we can safely rotate the ray of integration by the angle $\pi/2$, superposing it with imaginary axis. In this way, the integration contour passes by the aforementioned point $u=i$ from the right side. Accordingly, we have
\begin{eqnarray}
f_{ferm}(x)=-2\,{\rm Im}\,\bigg\{
\pi \, {\rm Res}_{u=i}\,F(u,x)\nonumber\\+ \,{\rm V.p.}\,\int\limits_0^\infty d\tilde{u}\, F(i\tilde{u},x)\bigg\}.\nonumber
\end{eqnarray}
Obviously, ${\rm Res}_{u=i}\,F(u,x)=-(i/2)e^{-\pi x/4}$. Since the function $F(i\tilde{u},x)$ is real valued for $0<\tilde{u}<1$, we can replace
the second term in the brackets simply by $\int_1^\infty d\tilde{u} F(i\tilde{u},x)$. For $\tilde{u}>1$, we can write $g(i\tilde{u})=g(i)-\int_1^{\tilde{u}}\sqrt{\tilde{v}^2-1}\,d\tilde{v}$, where $g(i)=i\pi/4$. Thus, after shifting the integration variables as $\tilde{u}\to\tilde{u}+1$, $\tilde{v}\to\tilde{v}+1$, we obtain
\begin{eqnarray}
\int\limits_1^\infty d\tilde{u} F(i\tilde{u},x)=-e^{-\pi x/4}\int\limits_0^\infty \frac{d\tilde{u}}{2\tilde{u}+\tilde{u}^2}\nonumber\\\times
\exp\left(-ix\int\limits_0^{\tilde{u}}\sqrt{2\tilde{v}+\tilde{v}^2}\,
d\tilde{v}\right).\nonumber
\end{eqnarray}
If $x$ is large as assumed, then the main contribution to the latter integral comes from the region of small $\tilde{u}$. Hence in the first approximation we can neglect the squares of $\tilde{u}$ and $\tilde{v}$. Finally, by transition to a new integration variable $\zeta=\int_0^{\tilde{u}}\sqrt{2\tilde{v}}\,d\tilde{v}
=(2\sqrt{2}/3){\tilde{u}}^{3/2}$, we are coming to
$$
{\rm Im}\,\int\limits_1^\infty d\tilde{u} F(i\tilde{u},x)=\frac13 e^{-\pi x/4}\int\limits_0^\infty d\zeta\,\frac{\sin(x\zeta)}{\zeta}=\frac{\pi}{6}e^{-\pi x/4}.
$$
Hence, we have $f_{ferm}(x)=(2\pi/3)e^{-\pi x/4}$. In principle, in this way the leading corrections of the order of $O(x^{-2/3}e^{-\pi x/4})$ can be derived as well. In the expression for $f_{scal}(x)$ the integrand differs only by the additional factor $u$. However, as it follows from our derivation, in the leading approximation the non-singular part of preexponent is all the same evaluated at $u=i$. Thus, $f_{scal}(x)=(2\pi i/3)e^{-\pi x/4}$.


\end{document}